\documentclass[aps,pra,twocolumn,showpacs]{revtex4}

\usepackage{amssymb}
\usepackage{amsmath}
\usepackage{graphicx}
\usepackage{dcolumn}
\usepackage{color}
\usepackage{bm}

\begin{document}
\newcommand{\eexp}[1]{\mbox{e}^{#1}}
\title{Manipulation of quantum particles in rapidly oscillating potentials by inducing phase hops}
\author{Armin Ridinger}
\email{ridinger@ens.fr}
\author{Christoph Weiss}
\email{weiss@theorie.physik.uni-oldenburg.de}

\affiliation{Laboratoire Kastler Brossel, \'Ecole Normale Sup\'erieure, Universit\'e Pierre et Marie-Curie-Paris 6, 24 rue Lhomond, CNRS, F-75231 Paris Cedex 05, France}
\date{\today}

%
\begin{abstract}
Analytical calculations show that the mean-motion of a quantum particle trapped
by a rapidly oscillating potential can be significantly manipulated by inducing phase hops,
i.e., by instantaneously changing the potential's phase. A phase hop can be visualized as
being the result of a collision with an imaginary particle which can be
controlled. Several phase hops can have accumulating effects on the particle's mean-motion, even if 
they transform the particle's Hamiltonian into its initial one. The theoretical predictions are verified by numerical simulations for
the one-dimensional Paul-trap. 
\end{abstract}
\pacs{37.10.Ty, 
37.10.Gh, 
03.65.Ge 
03.75.Ss 
} \maketitle
%

\section{Introduction}

Potentials which oscillate rapidly relative to the motion of particles inside them are widely
used to trap charged and neutral particles. 
Most notably, this is because rapidly oscillating potentials (ROPs) allow trapping in cases where static
potentials 
cannot. Well-known paradigms are the
Paul-trap for charged particles~\nocite{LeiBla03}\cite{LeiBla03,MajGhe04} and the electro-
and magneto-dynamic traps for high-field seeking polar
molecules~\nocite{JunRie04}\cite{JunRie04,VelBet05} and neutral
atoms~\nocite{CorMon91,KisHac06}\cite{CorMon91,KisHac06,SchMar07}. Furthermore, ROPs allow
the realization of complicated trap geometries. Prime examples are the
TOP-trap~\cite{PetAnd95,FraZam04}, the optical billiard
traps~\nocite{MilHan01}\cite{MilHan01,FriKap01} and rapidly scanning optical
tweezers~\nocite{AhmTim05}\cite{AhmTim05,SchOoi08} for ultracold neutral atoms or even microparticles such as polymers and
cells~\nocite{SasKos91}\cite{SasKos91,NeuBlo04}. Another reason is that the
description of the motion of particles in a ROP --- as compared to other time-varying
potentials --- is very simple: the particles' \textit{mean}-motion (averaged over the ROP's fast oscillations) is 
to a good approximation determined by a static \textit{effective} potential~\nocite{LanLif76,CooSha85,GroRak88}\cite{LanLif76,CooSha85,GroRak88,RahGil03}.

Preliminary calculations for the classical regime show that in ROPs with a vanishing time-average, as
e.g.~the Paul-trap, the mean-motion of trapped particles is strongly coupled to the phase of
the ROP~\cite{RidDav07}. Consequently, the particles' mean-motion can be appreciably
manipulated by changing the phase of the ROP.
For the Paul-trap, a phase hop can change the mean-energy of a trapped classical particle (that is not constantly
at rest) by a factor which can take any value between $0.1$ and $9.9$, independent of the
particle's mean-energy~\cite{RidDav07}, thus offering a powerful tool for particle
manipulation.  

However, often \textit{quantum} particles are trapped in
ROPs~\cite{LeiBla03,MajGhe04,PetAnd95,FraZam04,MilHan01,FriKap01,AhmTim05,SchOoi08}.
It is not clear if this tool would work for quantum particles: in the Paul-trap, a classical particle 
which does not move is not affected by a phase hop~\cite{RidDav07}. Thus, the
same might be true for a quantum particle in, e.g., the
ground state of the effective trapping potential.

In this article, we derive an independent quantum mechanical treatment of the effect of 
phase hops on a particle trapped by a ROP of arbitrary shape.
By both analytical and numerical calculations we show that a phase hop can strongly influence the
particle's mean-motion, even if it is in the ground state of the effective trapping
potential. The experimental ability to 
prepare single~\cite{LeiBla03,MajGhe04} 
and ensembles~\cite{PetAnd95,FraZam04,SchOoi08} of quantum particles in ROPs in specific states 
would allow to apply this tool in a controlled fashion.

The model used to describe both the time-dependent and the
effective system is given in Sec.~\ref{sec:qua}.
In Sec.~\ref{sec:eff} it is shown that the effect of a phase hop can be visualized as being the result of a collision with 
an imaginary particle. In Sec.~\ref{sec:pha} it is demonstrated 
that phase hops offer a powerful tool to manipulate quantum particles, whose application, in particular,
does not affect the effective trapping potential. 

\section{\label{sec:qua}Quantum motion in a rapidly oscillating potential}

\begin{figure}
\centering
\includegraphics[width=\linewidth]{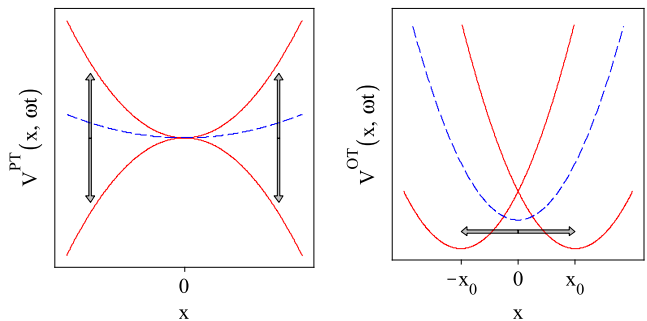}
\caption{(Color online) Model-potentials for the Paul-trap (\ref{PT}) (left) and for spatially oscillating optical tweezers (\ref{OT}) (right). Solid (red) curves: time-dependent potential, dashed (blue) curves: time-independent effective potential.}\label{Fig1}
\end{figure}

The Schr\"{o}dinger equation for a quantum particle in a time-periodic potential $V(x,\omega
t)$ reads
\begin{eqnarray}\label{se}
i\hbar\frac{\partial}{\partial t}\psi(x,t)=\hat{H}(x,\omega t)\psi(x,t),
\end{eqnarray}
with the time-dependent, periodic Hamiltonian
\begin{eqnarray}\label{Hamiltoniantimedependent}
\hat{H}(x,\omega t)=-\frac{\hbar^2}{2m}\frac{\partial^2}{\partial x^2}+V_0(x)+V_1(x,\omega t),
\end{eqnarray}
where the last two terms represent a separation of~$V$ into a time-averaged
part~$V_0$ and an oscillating part~$V_1$ with a vanishing period-average. Two experimentally relevant examples for the considered type of potentials are
\begin{eqnarray}\label{PT}
V^{\textrm{\scriptsize PT}}(x,\omega t)&=&\frac{1}{2}m\omega_{\textrm{\scriptsize osc}}^2x^2\cos\omega t,\\
\label{OT}
V^{\textrm{\scriptsize OT}}(x,\omega t)&=&\frac{1}{2}m\omega_{\textrm{\scriptsize osc}}^2(x-x_0\cos\omega t)^2,
\end{eqnarray}
(shown in Fig.~\ref{Fig1})
%
%
%
%
for which $V^{\textrm{\scriptsize PT}}_0(x)\!\equiv\!\overline{V^{\textrm{\scriptsize PT}}(x,\omega
  t)}\!=\!0$ (the overbar denotes the time-average over one period) and $V^{\textrm{\scriptsize OT}}_0(x)\!\equiv\!\overline{V^{\textrm{\scriptsize OT}}(x,\omega t)}\!=\!m\omega_{\textrm{\scriptsize osc}}^2(x^2\!-\!x_0^2/2)/2$. These potentials are model-potentials for the Paul-trap~(\ref{PT}) and for rapidly scanning optical tweezers~(\ref{OT}), respectively. If the potential's driving frequency~$\omega$ is sufficiently large, the particle's motion is separable into two parts that evolve on the different time scales $t$ and $\tau\!\equiv\!\omega t$~\nocite{CooSha85,GroRak88}\cite{CooSha85,GroRak88,RahGil03}. Floquet's theorem suggests that therefore the solutions of Eq.~(\ref{se}) approximately have the functional form~\nocite{CooSha85,GroRak88}\cite{CooSha85,GroRak88,RahGil03}
\begin{eqnarray}\label{Ansatz}
\psi(x,\tau,t)\approx e^{-iF(x,\tau)}\phi(x,t),
\end{eqnarray}
where $F$ is a time-periodic function~\cite{footnoteARCW1} with the same period as~$V$,
and~$\phi$ is a slowly varying function of time, which is the solution of a Schr\"{o}dinger
equation with a time-\textit{independent} Hamiltonian and which describes the particle's
mean-motion. The Schr\"{o}dinger equation for~$\phi$ is obtained by substituting
Eq.~(\ref{Ansatz}) into Eq.~(\ref{se}), choosing
\begin{eqnarray}\label{F}
F(x,\tau)\equiv\frac{1}{\hbar\omega}\!\left(\int_0^\tau\! V_1(x,\tau')d\tau'-\overline{\int_0^{\tau}\!V_1(x,\tau')d\tau'}\right),
\end{eqnarray}
(which implies~$\overline{F(x,\tau)}=0$) and averaging the resulting equation in time over one period of~$V$. This results in
\begin{eqnarray}\label{effse}
i\hbar\frac{\partial}{\partial t}\phi(x,t)=\hat{H}_{\textrm{\scriptsize eff}}(x)\phi(x,t),
\end{eqnarray}
with the time-independent \textit{effective} Hamiltonian
\begin{eqnarray}\label{Heff}
\hat{H}_{\textrm{\scriptsize eff}}(x)=-\frac{\hbar^2}{2m}\frac{\partial^2}{\partial x^2}+V_0(x)+\frac{\hbar^2}{2m}\overline{F'(x,\tau)^2}
\end{eqnarray}
(primes denote derivatives with respect to $x$). The last two terms of Eq.~(\ref{Heff}) represent a time-independent effective potential~$V_{\textrm{\scriptsize
    eff}}(x)\!\equiv\!V_0(x)\!+\!\frac{\hbar^2}{2m}\overline{F'(x,\tau)^2}$. For the above examples one has
$V_{\textrm{\scriptsize eff}}^{\textrm{\scriptsize PT}}(x)\!=\!m\Omega^2x^2/2$, with
$\Omega\!\equiv\!\omega_{\textrm{\scriptsize osc}}^2/(\omega\sqrt{2})$, and
$V_{\textrm{\scriptsize eff}}^{\textrm{\scriptsize
    OT}}(x)\!=\!m\omega_{\textrm{\scriptsize osc}}^2x^2/2+c$, with a
constant $c$ (see Fig.~\ref{Fig1}). The solution of Eq.~(\ref{effse}) is 
\begin{eqnarray}\label{phi}
        \phi(x,t)=e^{-i\,\hat{H}_{\textrm{eff}}(x)t/\hbar}e^{iF(x,0)}\psi(x,0),
\end{eqnarray}
where $\psi(x,0)$ is the particle's state at $t\!=\!0$. $\phi$ approximately describes the
particle's mean-motion, since it is $\overline{\langle x\rangle}\!\equiv\!\overline{\langle\psi|\hat{x}|\psi\rangle}\!\approx\!\langle\phi|\hat{x}|\phi\rangle$ and $\overline{\langle p\rangle}\!\equiv\!\overline{\langle\psi|\hat{p}|\psi\rangle}\!\approx\!\langle\phi|\hat{p}|\phi\rangle$. The difference between $\phi$ and $\psi$ is approximately given by the oscillating phase factor $e^{-iF}$ (Eq.~(\ref{Ansatz})), which has a small amplitude as $F$ scales with~$\omega^{-1}$ (Eq.~(\ref{F})), and thus describes a \textit{micro}motion~\cite{LeiBla03,MajGhe04}, since it is $\langle\psi|\hat{x}|\psi\rangle\!-\!\langle\phi|\hat{x}|\phi\rangle\!\approx\!0$ and $\langle\psi|\hat{p}|\psi\rangle\!-\!\langle\phi|\hat{p}|\phi\rangle\!\approx\!-\hbar\langle\phi|F'|\phi\rangle$. In the following we consider the limit $\omega\rightarrow\infty$, where the approximations become exact. We consider the case of a trapped particle, which can be expressed in terms of eigenstates of the effective Hamiltonian (the \textit{stationary mean-motion states}), i.e., for the above examples (Eqs.~(\ref{PT}) and (\ref{OT})), of the harmonic oscillator.

\section{\label{sec:eff}Effects of phase hops: collisions with imaginary particles}

Suppose now that, at a time $t_{\textrm{\scriptsize ph}}$, the phase of the potential
$V(x,\tau)$ is instantaneously changed from $\tau_{\textrm{\scriptsize ph}}\!\equiv\!\omega
t_{\textrm{\scriptsize ph}}$ to $\tau_{\textrm{\scriptsize ph}}\!+\!\Delta\varphi$. Then, for
$t\!>\!t_{\textrm{\scriptsize ph}}$ the particle is moving in the ROP
$V_{\textrm{new}}(x,\tau)\!\equiv\!V(x,\tau\!+\!\Delta\varphi)$ and its
mean-motion wave function $\phi_{\textrm{\scriptsize new}}$ is governed by
Eq.~(\ref{effse}) with an effective Hamiltonian $\hat{H}^{\textrm{new}}_{\textrm{eff}}$. As the
effective Hamiltonian of a ROP consists only of period-averaged terms~(Eq.~(\ref{Heff})), it is independent of the phase of the ROP, implying $\hat{H}^{\textrm{new}}_{\textrm{eff}}=\hat{H}_{\textrm{eff}}$. 
Thus the equation of motion for the particle's mean-motion wave function remains unchanged. 
However, the mean-motion wave function itself changes due to the natural
continuity of the particle's real wave function at $t\!=\!t_{\textrm{\scriptsize ph}}$: For $t\!\neq\!t_{\textrm{\scriptsize ph}}$ the latter is a product of a phase factor and the mean-motion wave function (Eq.~(\ref{Ansatz})). As the phase factor ($e^{-iF}$) depends on the phase of~$V$ through $F$~(Eq.~(\ref{F})), it changes instantaneously and thus involves a corresponding change of the mean-motion wave function. In the first instance one might, however, naively expect that this change is negligible since for large~$\omega$ the change of the phase factor is very small ($F$ scales with $\omega^{-1}$). But, as we show in the following, the change of the particle's mean-motion can be indeed significant and even for arbitrarily large $\omega$. 

To demonstrate this, we calculate $\phi_{\textrm{\scriptsize new}}$ and derive the resulting changes of the mean-motion
observables. The condition of continuity for the particle's real wave function yields
\begin{eqnarray}\label{phinewtph}
        \phi_{\textrm{\scriptsize new}}(x,t_{\textrm{\scriptsize ph}})=e^{i\Delta F(x,\tau_{\textrm{\tiny ph}})}\phi(x,t_{\textrm{\scriptsize ph}})
\end{eqnarray}
(the notation $\lim_{t\rightarrow t_{\textrm{ph}},t<t_{\textrm{ph}}}\phi(x,t)\equiv\phi(x,t_{\textrm{\scriptsize ph}})$ and $\lim_{t\rightarrow t_{\textrm{ph}},t>t_{\textrm{ph}}}\phi_{\textrm{\scriptsize new}}(x,t)\equiv\phi_{\textrm{\scriptsize new}}(x,t_{\textrm{\scriptsize ph}})$ is used),
where $\Delta F(x,\tau_{\textrm{\scriptsize ph}})\!\equiv\! F(x,\tau_{\textrm{\scriptsize ph}}\!+\!\Delta\varphi)\!-\!F(x,\tau_{\textrm{\scriptsize ph}})=\frac{1}{\hbar\omega}\!\int_0^{\tau_{\textrm{\tiny ph}}}\!\left[V_1(x,\tau\!+\!\Delta\varphi)-V_1(x,\tau)\right]d\tau$.
Applying Eq.~(\ref{effse}) leads to
\begin{eqnarray}\label{phinew}
        \phi_{\textrm{\scriptsize new}}(x,t)=e^{-i\,(t-t_{\textrm{\tiny ph}})\hat{H}_{\textrm{eff}}(x)/\hbar}e^{i\Delta F(x,\tau_{\textrm{\tiny ph}})}\phi(x,t_{\textrm{\scriptsize ph}}).
\end{eqnarray}
Combined with Eq.~(\ref{phi}), Eq.~(\ref{phinew}) allows to completely describe the phase
hop, as the particle's mean-motion wave function is known for all times.

The effect of the phase hop on $\phi$ involves an instantaneous change of some mean-motion observables, which, for a given mean-motion observable $\hat{O}$, is given by $\Delta \overline{\langle O\rangle}\!=\!\langle \phi_{\textrm{\scriptsize new}}(x,t_{\textrm{\scriptsize ph}})|\hat{O}|\phi_{\textrm{\scriptsize new}}(x,t_{\textrm{\scriptsize ph}})\rangle\!-\!\langle \phi(x,t_{\textrm{\scriptsize ph}})|\hat{O}|\phi(x,t_{\textrm{\scriptsize ph}})\rangle$. Using Eqs.~(\ref{phi}) and~(\ref{phinewtph}) we find
\begin{eqnarray}\label{Deltaxgeneral}
\Delta \overline{\langle x\rangle}&=&0,\\
\label{Deltapgeneral}
\Delta \overline{\langle p\rangle}&=&\hbar\int_{-\infty}^{\infty}|\phi(x,t_{\textrm{\scriptsize ph}})|^2\Delta F'(x,\tau_{\textrm{\scriptsize ph}})\,dx.
\end{eqnarray}
Inspection of the right hand side (rhs) of Eq.~(\ref{Deltapgeneral}) shows that $\Delta \overline{\langle p\rangle}$ equals the change of momentum of the particle's micromotion, which is taking place at the same time, demonstrating that the phase hop causes a momentum transfer between the micromotion and the mean-motion. The fact that the phase hop can change the particle's mean-momentum instantaneously, but not its mean-position, shows that its effect on the particle's mean-motion can be visualized as being the result of a collision with an imaginary particle~\cite{footnoteARCW2}.  The particle's mean-energy, which is conserved before and after the phase hop, changes by
\begin{eqnarray}\label{DeltaEgeneral}
\Delta \overline{\langle E\rangle}&=&\frac{\hbar^2}{2m}i\int_{-\infty}^{\infty}\phi'(x,t_{\textrm{\scriptsize ph}})^*\phi(x,t_{\textrm{\scriptsize ph}})\Delta    
     F'(x,\tau_{\textrm{\scriptsize ph}})\,dx\nonumber\\
  & &-\frac{\hbar^2}{2m}i\int_{-\infty}^{\infty}\phi(x,t_{\textrm{\scriptsize ph}})^*\phi'(x,t_{\textrm{\scriptsize ph}})\Delta 
     F'(x,\tau_{\textrm{\scriptsize ph}})\,dx\nonumber\\
  & &+\frac{\hbar^2}{2m}\int_{-\infty}^{\infty}|\phi(x,t_{\textrm{\scriptsize ph}})|^2(\Delta F'(x,\tau_{\textrm{\scriptsize ph}}))^2\,dx
\end{eqnarray}
(the notation $\lim_{t\rightarrow t_{\textrm{ph}},t<t_{\textrm{ph}}}\phi'(x,t)\equiv\phi'(x,t_{\textrm{\scriptsize ph}})$ is used).
Inspection of the rhs of Eq.~(\ref{DeltaEgeneral}) shows that $\Delta \overline{\langle
  E\rangle}$ is always non-negative if the particle is in a stationary mean-motion state $\phi_n$ (i.e.,~its mean-motion is in an eigenstate  $\phi_n$ of $\hat{H}_{\textrm{eff}}$):
\begin{equation}
\label{eq:ineq}
\Delta \overline{\langle  E_n\rangle} \ge 0\;,
\end{equation}
since it is $\phi_n'^*\phi_n=\phi_n^*\phi_n'$~\cite{footnoteARCW5}. Using the picture of the imaginary collision, Eq.~(\ref{eq:ineq}) also directly follows from the fact that stationary mean-motion states have $\overline{\langle p\rangle}=0$. For the Paul trap potential~(\ref{PT}) the fact that the energy change $\Delta \overline{\langle  E_n\rangle}$ can be non-zero (and even be very large as shown in Sec.~\ref{sec:pha}) marks a difference to the classical regime, since for a classical particle whose mean-motion is at rest, $\Delta \overline{E}_{\textrm{\scriptsize class.}}$ is always zero~\cite{RidDav07}. This difference is a direct consequence of Heisenberg's uncertainty principle which implies
that in the quantum regime the particle's mean-position is spread around zero, leading --- contrarily to
the classical regime --- to a non-vanishing micromotion (since $e^{-iF}\not\equiv1$ for
$x\neq0$), thus giving rise to an effect of the phase hop. If the quantum particle is not in a stationary mean-motion state its mean-energy
can due to Eq.~(\ref{DeltaEgeneral}) be both increased and decreased:
\begin{equation}
\label{Eposneg}
 \Delta \overline{\langle  E\rangle} \gtreqqless 0
\end{equation}
(see also the appendix).

\section{\label{sec:pha}Phase hops can have significant impact}

In order to demonstrate that the phase hop can have a strong effect we compare $\Delta
\overline{\langle E\rangle}$ to the particle's 
initial mean-energy $\overline{\langle E\rangle}$. Experimentally relevant are particles that are in 
stationary mean-motion states $\phi_n$ with mean-energy $E_n$. 
For the Paul-trap~(\ref{PT}) we find
\begin{eqnarray}\label{DeltaEnPTabsolut}
  \Delta \overline{\langle E_n^{\textrm{\scriptsize PT}}\rangle}=[\sin(\tau_{\textrm{\scriptsize ph}}\!+\!\Delta\varphi)-\sin(\tau_{\textrm{\scriptsize ph}})]^2\,E_n^{\textrm{\scriptsize PT}},
\end{eqnarray}
where~$E_n^{\textrm{\scriptsize PT}}\!=\!\hbar\Omega(n\!+\!1/2)$. Thus, the relative change $\Delta \overline{\langle E_n^{\textrm{\scriptsize PT}}\rangle}/E_n^{\textrm{\scriptsize PT}}$ is independent of~$n$ and of~$\omega$, and it can take values between~0 and~4. This demonstrates that in the Paul-trap a phase hop can have a strong effect on the particle's mean-motion, even for arbitrarily large~$\omega$. For the rapidly scanning optical tweezers~(\ref{OT}) we find 
\begin{eqnarray}\label{DeltaEnOTabsolut}
  \Delta \overline{\langle E_n^{\textrm{\scriptsize OT}}\rangle}=\frac{\omega_{\textrm{\scriptsize ref}}^2}{\omega^2}\,[\sin(\tau_{\textrm{\scriptsize ph}}\!+\!\Delta\varphi)-\sin(\tau_{\textrm{\scriptsize ph}})]^2\,E_0^{\textrm{\scriptsize OT}},
\end{eqnarray}
with~$E_n^{\textrm{\scriptsize OT}}\!=\!\hbar\omega_{\textrm{\scriptsize osc}}(n\!+\!1/2)$
and the reference frequency $\omega_{\textrm{\scriptsize
    ref}}\!\equiv\!\sqrt{m\omega_{\textrm{\scriptsize osc}}^3x_0^2/\hbar}$. Here, the
relative change $\Delta \overline{\langle E_n^{\textrm{\scriptsize
      OT}}\rangle}/E_n^{\textrm{\scriptsize OT}}$ can
take values between 0 and $[4/(2n\!+\!1)]\,\omega_{\textrm{\scriptsize ref}}^2/\omega^2$ and thus becomes negligible for $\omega\!\rightarrow\!\infty$. 
However, as an inspection of $\omega_{\textrm{\scriptsize
    ref}}$ shows, $\Delta \overline{\langle E_n^{\textrm{\scriptsize OT}}\rangle}/E_n^{\textrm{\scriptsize OT}}$ can still be large even if $\omega$ is as large as required by the validity condition of the underlying effective theory (i.e.~for~$\omega\!\gg\!\omega_{\textrm{\scriptsize osc}}$, cf. Sec.~\ref{sec:qua}).

To generalize the above findings, consider first an arbitrary ROP with a vanishing time-average. The mean potential energy of a particle in a stationary mean-motion state $\phi_n$ then is $E_n^{\textrm{pot}}\!=\!\langle \phi_n|V_{\textrm{\scriptsize eff}}|\phi_n\rangle$
with $V_{\textrm{\scriptsize eff}}(x)\!=\!\frac{\hbar^2}{2m}\overline{F'(x,\tau)^2}$ and the following relation holds:
\begin{eqnarray}\label{DeltaEPotentialmeanzero}
        E_n^{\textrm{pot}}&<& \max_{\tau_{\textrm{\tiny ph}}}\left[\langle \phi_n(x)|\mbox{$\frac{\hbar^2}{2m}$}
                              F'(x,\tau_{\textrm{\scriptsize ph}})^2|\phi_n(x)\rangle\right]\nonumber\\
                          & &<\ \max_{\tau_{\textrm{\tiny ph}},\Delta \varphi}\left[\langle \phi_n(x)|
                               \mbox{$\frac{\hbar^2}{2m}$}(\Delta F'(x,\tau_{\textrm{\scriptsize ph}}))^2|\phi_n(x)\rangle\right]\nonumber\\
                          & &\qquad=\  \max_{\tau_{\textrm{\tiny ph}},\Delta \varphi}\left[\Delta \overline{\langle E_n\rangle}\right].
\end{eqnarray}
Therefore a time $\tau_{\textrm{\scriptsize ph}}$ exists (within each period of $V$) for which a phase hop of a size $\Delta\varphi$ (with $0\!\leq\!\Delta\varphi\!<\!2\pi$) induces a change $\Delta \overline{\langle E_n\rangle}$ of the particle's mean-energy $E_n$ which is greater than its mean potential energy $E_n^{\textrm{pot}}$. Since $E_n^{\textrm{pot}}$ is in general a significant fraction of $E_n$, Eq.~(\ref{DeltaEPotentialmeanzero}) shows that the phase hop can \textit{always} be induced such that $\Delta \overline{\langle E_n\rangle}$ is large with respect to $E_n$, even for arbitrarily large $\omega$. An intuitive explanation of the very fact that in ROPs with a vanishing time-average a phase hop can always significantly change the particle's mean-energy (when it is induced in a correct moment), can be given as follows: For ROPs with a vanishing time-average the particle's mean potential energy equals the average kinetic energy that is stored in the particle's micromotion (since $\langle\phi|V_{\textrm{\scriptsize eff}}|\phi\rangle=(\overline{\langle\psi|\hat{p}^2|\psi\rangle-\langle\phi|\hat{p}^2|\phi\rangle})/(2m)$). A phase hop causes a momentum transfer between the particle's micromotion and mean-motion, whose maximum value is given by the peak-to-peak amplitude of the (oscillating) momentum of the micromotion (Eq.~(\ref{Deltapgeneral})). Since the momentum of the particle's micromotion is oscillating around \textit{zero}, this momentum transfer can, due to the equivalence of the micromotion's average kinetic energy and the mean potential energy, lead to a change of the particle's mean-energy which is comparable to its mean potential energy and which thus is significant. 

In ROPs with a non-vanishing time-average a phase hop can only then lead to a significant change of the particle's mean-energy if~$\omega$ is not too large, since for such ROPs the fraction of the particle's mean potential energy which equals the average kinetic energy stored in the particle's micromotion scales with~$\omega^{-2}$. This is expressed by Eq.~(\ref{DeltaEgeneral}), which yields that for stationary mean-motion states $\Delta \overline{\langle E_n\rangle}$ scales with $\omega^{-2}$ (cf.~Eq.~(\ref{DeltaEnOTabsolut})).

\begin{figure}
\centering
\includegraphics[width=\linewidth]{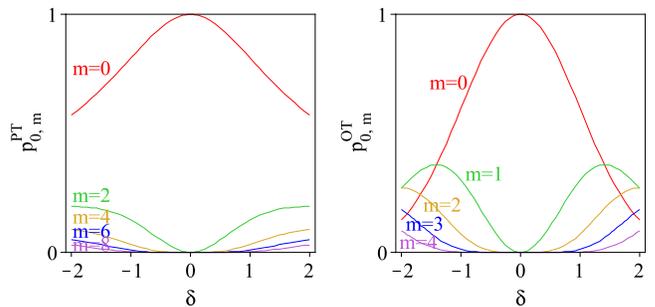}
\caption{(Color online) The experimentally measurable~\cite{LeiBla03,MajGhe04} probabilities for the transitions between the mean-motion ground state 
                        $\phi_0$ and the mean-motion states $\phi_m$ induced by a phase hop 
                        for a particle in the Paul-trap~(\ref{TransProbPT}) (left) and the rapidly 
                        scanning optical tweezers~(\ref{TransProbOT}) 
                        (right, with $\omega\!=\!\omega_{\textrm{\scriptsize ref}}$) as a function
                        of the parameter $\delta\!=\!\sin(\tau_{\textrm{\scriptsize
    ph}}\!+\!\Delta\varphi)\!-\!\sin(\tau_{\textrm{\scriptsize ph}})$.}\label{Fig2}
\end{figure}

We have seen that the phase hop can affect the particle's mean-motion wave function and observables.
The phase hop thus can induce transitions between stationary mean-motion states, $\phi_n\!\rightarrow\!\phi_m$,  
whose probabilities are given by
$p_{n,m}\!\equiv\!|\langle\phi_{(n)}^{\textrm{\scriptsize
    new}}(x,t_{\textrm{ph}})|\phi_m(x)\rangle|^2$ (implying $p_{n,m}\!=\!p_{m,n}$), where $\phi^{\textrm{\scriptsize
    new}}_{(n)}$ denotes the particle's mean-motion state after the phase hop. For the
potentials~(\ref{PT}) and~(\ref{OT}), the $p_{n,m}$ can be calculated analytically. In particular, the $p_{0,m}$ are of experimental relevance as the mean-motion ground state $\phi_0$ can be prepared with
a high precision and can be easily probed. For the Paul-trap~(\ref{PT}) we find 
\begin{eqnarray}\label{TransProbPT}
        & &\!\!\!\!\!\!p^{\textrm{\scriptsize PT}}_{0,m}=\left(\!\frac{m!}{2^{\frac{3m}{2}}\left(\frac{m}{2}\right)!\,\left(\frac{m}{2}\right)!}\!\right)\!\frac{\delta^m}{\left(1+\frac{\delta^2}{2}\right)^{\frac{m+1}{2}}},\ \textrm{ for }m=\textrm{even},\nonumber\\
        & &\!\!\!\!\!\!p^{\textrm{\scriptsize PT}}_{0,m}=0,\ \textrm{ for } m=\textrm{odd},
\end{eqnarray}
with $\delta\!=\!\sin(\tau_{\textrm{\scriptsize
    ph}}\!+\!\Delta\varphi)\!-\!\sin(\tau_{\textrm{\scriptsize ph}})$. 
In typical single ion experiments, the $p_{0,m}$ could be directly measured using resolved Raman sideband spectroscopy~\cite{LeiBla03,MajGhe04}. Figure~{\ref{Fig2}} shows that the probability for a particle to remain in the mean-motion ground state can be as small as $58\%$, demonstrating the significance of the effect of the phase hop. For the rapidly scanning optical tweezers~(\ref{OT}) we find 
\begin{eqnarray}\label{TransProbOT}
p^{\textrm{\scriptsize OT}}_{0,m}&=&\frac{1}{2^m\,m!}\left(\frac{\omega_{\textrm{ref}}}{\omega}\,\delta\right)^{2m}e^{-\frac{1}{2}\left(\frac{\omega_{\textrm{ref}}}{\omega}\delta\right)^2}.
\end{eqnarray}
As $p^{\textrm{\scriptsize OT}}_{0,0}\!\rightarrow\!1$ for~$\omega\!\rightarrow\!\infty$, the
effect of the phase hop becomes negligible for too large $\omega$. However, Fig.~\ref{Fig2} shows 
that for $\omega\!=\!\omega_{\textrm{\scriptsize ref}}$ the effect of the phase hop is still 
significant. For weakly interacting bosonic quantum gases, $p^{\textrm{\scriptsize
    OT}}_{0,0}$ could be determined by measuring the number of atoms that remain in a
Bose-Einstein condensate~\cite{SchOoi08}, provided that the measurement is performed immediately
  after the phase hop before a rethermalisation takes place. For an atomic gas of degenerate fermions, phase hops
(cf. Eq.~(\ref{DeltaEnOTabsolut})) could offer a tool to more quickly increase
the energy and thus, after thermalization, the temperature in a controlled way~\cite{footnoteARCW4} without having to change or to switch off-and-back-on the trapping potential and to in-between await an expansion of the gas~\cite{ThoKin05}.

%
%

\begin{figure}
\centering
\includegraphics[angle=270,width=\linewidth]{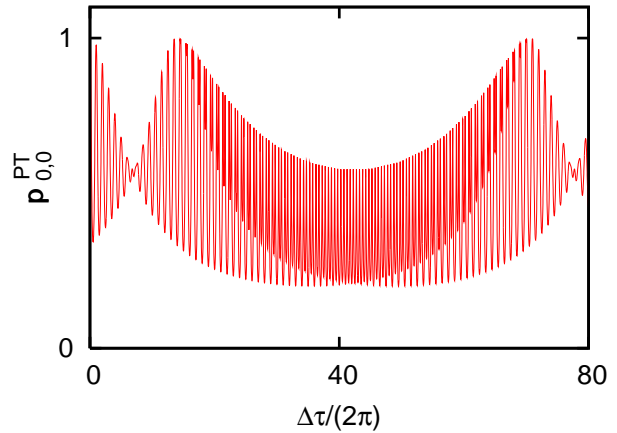}
\caption{(Color online) The experimentally measurable~\cite{LeiBla03,MajGhe04} probability for a particle in the Paul-trap~(\ref{PT}) to return
                        to the mean-motion ground state $\phi_0$ when two phase hops of size 
                        $\Delta\varphi\!=\!\pi$ are successively induced (at times 
                        $\tau_{\textrm{ph}}^{(1)}\!=\!\pi/2$ and 
                        $\tau_{\textrm{ph}}^{(2)}\!=\!\tau_{\textrm{ph}}^{(1)}\!+\!\Delta\tau$) 
                        as a function of $\Delta\tau$. Although one might naively expect those two 
                        phase-hops cancel each other, the second phase hop can, in fact, further reduce 
                        the probability to return to the ground state. Even parameters for which $p^{\textrm{\scriptsize PT}}_{0,0}=1$
                        are interesting experimentally: they can be used to verify if the
                        system indeed is described by dissipation-less quantum mechanics.}\label{Fig3}
\end{figure}

The manipulation by phase hops can be made more effective by inducing several phase hops
successively. Figure~\ref{Fig3} shows the transition probability $p^{\textrm{\scriptsize
    PT}}_{0,0}$ for two successively induced phase hops of size $\Delta\varphi\!=\!\pi$ as a function of
their time delay. Although the
effects of the two phase hops on the particle's ($2\pi$-periodic)
Hamiltonian~(\ref{Hamiltoniantimedependent}) cancel each other, their effects on the
particle's mean-motion do not necessarily cancel and can even be more significant as in the case of a single phase hop.

%
%

\begin{figure}
\centering
\includegraphics[angle=270,width=\linewidth]{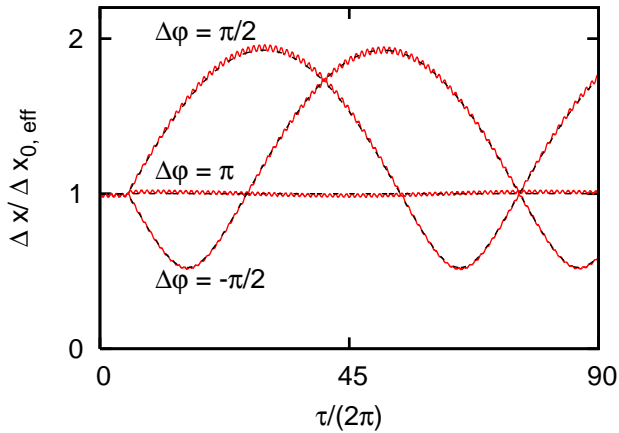}\\
\caption{(Color online) Time-evolution of the rms-deviation $\Delta x$ for a particle in the 
                        Paul-trap~(\ref{PT}) which initially is in the mean-motion ground 
                        state, affected by a phase hop of different sizes $\Delta\varphi$ 
                        (normalized to the theoretically predicted initial mean-motion value 
                        $\Delta x_{0,\rm eff}=\sqrt{\hbar/(2m\Omega)}$). Solid (red) curves: 
                        results of numerical simulations, dashed (black) curves: theoretical 
                        mean-motion prediction. Driving frequency: 
                        $\omega\!\approx\!70\,\Omega$.
}\label{Fig4}
\end{figure}

To countercheck our theoretical predictions, we performed numerical
simulations for a particle in the Paul-trap potential~(\ref{PT}) by integrating the full
time-dependent Schr\"{o}dinger equation~(\ref{se}). Figure~\ref{Fig4} shows
the time-evolution of the experimentally measurable root-mean-square (rms)-deviation $\Delta x\!\equiv\!\sqrt{\langle\psi|\hat{x}^2|\psi\rangle-\langle\psi|\hat{x}|\psi\rangle^2}$ of the particle's position when influenced by a phase hop.
The initial state was chosen to be $\psi(x,0)\!=\!e^{-iF^{\textrm{\tiny
      PT}}(x,0)}\phi_0^{\textrm{\scriptsize PT}}(x)$, which determines the particle to be in
the mean-motion ground state $\phi_0^{\textrm{\scriptsize PT}}$. 
Figure~\ref{Fig4} shows very good agreement between numerics and the
theoretical mean-motion predictions derived via computer-algebra from
Eqs.~(\ref{phi}) and~(\ref{phinew}) (the small oscillations of the solid (red) curves around their own mean represent higher orders of the particle's micromotion~\cite{footnoteARCW1}, which had been disregarded in the derivation of the theoretical mean-motion predictions (depicted as dashed (black) curves) and which disappear for larger $\omega$). The driving frequency used in the simulation is $\omega\!\approx\!70\,\Omega$, and thus the numerics demonstrate that the theoretical predictions for the limit $\omega\!\rightarrow\!\infty$ even hold for such small $\omega$. For larger $\omega$ the agreement between both approaches would even be better. Further, the numerics allowed to obtain a practical definition for ``instantaneous'': in experiments the phase-hop must happen on time-scales smaller than the period of $V$~\cite{footnoteARCW3}.

%
%

The results presented in this article must be consistent with the classical results of Ref.~\cite{RidDav07}. To countercheck this we calculated $\Delta \overline{\langle E\rangle}$ for the case that the mean-motion of a quantum particle in the Paul-trap potential is in a \textit{coherent} state~\cite{LeiBla03,MajGhe04}, and showed that its classical limit equals the corresponding classical result (see appendix). 

\section{Conclusion}

In this article we have presented a quantum mechanical treatment of the effect of phase hops in ROPs 
on a single trapped particle. We have computed the particle's mean-motion wave function
for all times and the resulting consequences on its mean-motion observables. We have calculated the transition probabilities between stationary mean-motion states and we have shown that the particle's mean-energy can in general be both increased and decreased, except if it is in a stationary mean-motion state, then the mean-energy can only be increased. In particular we have shown that the effect of a phase hop can be very strong.

Both for classical and for quantum particles, the induction
of phase hops provides a powerful tool for particle manipulation. Besides its strong effect it has the following further appealing properties: firstly, its experimental implementation would be very simple and would not even require a change of an existing setup. Secondly, its application would be a controlled operation since a phase hop does not have an influence on the effective trapping potential, because that is independent of the phase of the ROP. Finally, it could be applied to any kind of trappable particles, because its mechanism is simply based on changes of the ROP and does thus not rely on particular internal properties of the particles as do laser-based manipulation methods.

In this article we have restricted our considerations to a single particle. Our main goal was to obtain analytical results, as these directly apply to a widely used and studied experimental system, the single ion in a Paul-trap~\cite{LeiBla03,MajGhe04}. Due to the experimentally achievable pureness of this system, it can be used to precisely investigate phase hops and their possible applications experimentally. One of such possible applications could be to decelerate a single ion, which we here showed to be possible also in the quantum regime.

For future research, it will be interesting to study the effect of phase hops on ensembles of
(interacting) particles. The extension of single particle effects in
time-periodic systems~\cite{GrossmannEtAl91,KierigEtAl08} to multi-particle systems is an
active field of theoretical and experimental physics~\cite{EckardtEtAl05,SiasEtAl07}. An
experimental investigation of the schemes proposed in this manuscript could be performed with cold ion clouds stored in a Paul-trap~\cite{MajGhe04} or ultracold neutral atoms stored in a TOP-trap~\cite{PetAnd95,FraZam04} or in rapidly scanning optical tweezers~\nocite{AhmTim05}\cite{AhmTim05,SchOoi08}. In the field of degenerate Fermi gases, phase hops could offer a convenient way to realize controlled heating~\cite{footnoteARCW4}.

\acknowledgments

We thank D.~Gu\'ery-Odelin, N.~Davidson, R.~Ozeri, C.~Champenois, N.~Navon and F. Chevy for insightful
discussions and C.~Salomon for reading the manuscript. We acknowledge funding by the German
Federal Ministry of Education and Research (A.R.), the European Science Foundation (A.R) and
the European Union (C.W., contract MEIF-CT-2006-038407).

\begin{appendix}

\section*{Appendix: Effect of a phase hop on coherent mean-motion states}

Let the mean-motion of a quantum particle in the Paul-trap potential~(\ref{PT}) be in a \textit{coherent} state. This is a quantum state which describes a wave packet that follows the motion of a classical particle in a harmonic oscillator potential while retaining its shape (thus also referred to as \textit{quasiclassical} state). As the effective potential of the Paul-trap is harmonic, the particle's mean-motion possesses coherent states which are given by~\cite{LeiBla03,MajGhe04}
\begin{eqnarray}\label{CS}
\phi_{\textrm{\scriptsize CS}}(x)=\left(\frac{m\Omega}{\pi\hbar}\right)^{\frac{1}{4}}e^{\frac{i}{\hbar}\overline{\langle p\rangle} x}e^{-\frac{m\Omega}{2\hbar}(x-\overline{\langle x\rangle})^2},
\end{eqnarray}
where Eq.~(\ref{CS}) describes a quantum particle with mean-position~$\overline{\langle x\rangle}$, mean-momentum~$\overline{\langle p\rangle}$ and mean-energy $\overline{\langle E_{\textrm{\scriptsize CS}}\rangle}\!=\!\langle\phi_{\textrm{\scriptsize CS}}|\hat{p}^2|\phi_{\textrm{\scriptsize CS}}\rangle/(2m)\!+\!\langle\phi_{\textrm{\scriptsize CS}}|V_{\textrm{\scriptsize eff}}^{\textrm{\scriptsize PT}}(\hat{x})|\phi_{\textrm{\scriptsize CS}}\rangle$. When a phase hop is induced, $\overline{\langle E_{\textrm{\scriptsize CS}}\rangle}$ changes according to Eq.~(\ref{DeltaEgeneral}) by
\begin{eqnarray}\label{CoherentStates}
  \Delta \overline{\langle E_{\textrm{\scriptsize CS}}\rangle}=\sqrt{2}\,\Omega\overline{\langle x\rangle}\,\overline{\langle p\rangle}\delta + m\Omega^2\langle\phi_{\textrm{\scriptsize CS}}|\hat{x}^2|\phi_{\textrm{\scriptsize CS}}\rangle\delta^2,
\end{eqnarray}
where $\delta\!=\!\sin(\tau_{\textrm{\scriptsize
    ph}}\!+\!\Delta\varphi)\!-\!\sin(\tau_{\textrm{\scriptsize ph}})$. A classical particle with mean-position~$\overline{x}$, mean-momentum~$\overline{p}$ and mean-energy $\overline{E}_{\textrm{\scriptsize class.}}\!=\!\overline{p}^2/(2m)\!+\!V_{\textrm{\scriptsize eff}}^{\textrm{\scriptsize PT}}(\overline{x})$ would, in the case of the phase hop, change its mean-energy by $\Delta \overline{E}_{\textrm{\scriptsize class.}}\!=\!\sqrt{2}\,\Omega \overline{x}\,\overline{p}\,\delta\!+\!m\Omega^2\overline{x}^2\delta^2$~\cite{RidDav07}, which is the classical limit of Eq.~(\ref{CoherentStates}) (since it is $\lim_{\hbar\rightarrow0}\overline{\langle x\rangle}=\overline{x}$, $\lim_{\hbar\rightarrow0}\overline{\langle p\rangle}=\overline{p}$, $\lim_{\hbar\rightarrow0}\langle\phi_{\textrm{\scriptsize CS}}|\hat{x}^2|\phi_{\textrm{\scriptsize CS}}\rangle=\lim_{\hbar\rightarrow0}\langle\phi_{\textrm{\scriptsize CS}}|\hat{x}|\phi_{\textrm{\scriptsize CS}}\rangle^2=\overline{x}^2$ and $\lim_{\hbar\rightarrow0}\langle\phi_{\textrm{\scriptsize CS}}|\hat{p}^2|\phi_{\textrm{\scriptsize CS}}\rangle=\lim_{\hbar\rightarrow0}\langle\phi_{\textrm{\scriptsize CS}}|\hat{p}|\phi_{\textrm{\scriptsize CS}}\rangle^2=\overline{p}^2$). This demonstrates that the quantum mechanical results presented in this article are consistent with the classical results of Ref.~\cite{RidDav07}. In particular it shows that in the quantum regime as well as in the classical regime the particle's mean-energy can in general be both increased and decreased by inducing a phase hop (cf. Eq.~(\ref{Eposneg})).  

\end{appendix}

%

%


\begin{thebibliography}{31}
\expandafter\ifx\csname natexlab\endcsname\relax\def\natexlab#1{#1}\fi
\expandafter\ifx\csname bibnamefont\endcsname\relax
  \def\bibnamefont#1{#1}\fi
\expandafter\ifx\csname bibfnamefont\endcsname\relax
  \def\bibfnamefont#1{#1}\fi
\expandafter\ifx\csname citenamefont\endcsname\relax
  \def\citenamefont#1{#1}\fi
\expandafter\ifx\csname url\endcsname\relax
  \def\url#1{\texttt{#1}}\fi
\expandafter\ifx\csname urlprefix\endcsname\relax\def\urlprefix{URL }\fi
\providecommand{\bibinfo}[2]{#2}
\providecommand{\eprint}[2][]{\url{#2}}

\bibitem[{\citenamefont{Leibfried et~al.}(2003)\citenamefont{Leibfried, Blatt,
  Monroe, and Wineland}}]{LeiBla03}
\bibinfo{author}{\bibfnamefont{D.}~\bibnamefont{Leibfried}},
  \bibinfo{author}{\bibfnamefont{R.}~\bibnamefont{Blatt}},
  \bibinfo{author}{\bibfnamefont{C.}~\bibnamefont{Monroe}}, \bibnamefont{and}
  \bibinfo{author}{\bibfnamefont{D.}~\bibnamefont{Wineland}},
  \bibinfo{journal}{Rev. Mod. Phys.} \textbf{\bibinfo{volume}{75}},
  \bibinfo{pages}{281} (\bibinfo{year}{2003}).

\bibitem[{\citenamefont{Major et~al.}(2004)\citenamefont{Major, Gheorghe, and
  Werth}}]{MajGhe04}
\bibinfo{author}{\bibfnamefont{F.~G.} \bibnamefont{Major}},
  \bibinfo{author}{\bibfnamefont{V.~N.} \bibnamefont{Gheorghe}},
  \bibnamefont{and} \bibinfo{author}{\bibfnamefont{G.}~\bibnamefont{Werth}},
  \emph{\bibinfo{title}{Charged Particle Traps: Physics and Techniques of
  Charged Particle Field Confinement}} (\bibinfo{publisher}{Springer-Verlag},
  \bibinfo{address}{New York}, \bibinfo{year}{2004}).

\bibitem[{\citenamefont{Junglen et~al.}(2004)\citenamefont{Junglen, Rieger,
  Rangwala, Pinkse, and Rempe}}]{JunRie04}
\bibinfo{author}{\bibfnamefont{T.}~\bibnamefont{Junglen}},
  \bibinfo{author}{\bibfnamefont{T.}~\bibnamefont{Rieger}},
  \bibinfo{author}{\bibfnamefont{S.~A.} \bibnamefont{Rangwala}},
  \bibinfo{author}{\bibfnamefont{P.~W.~H.} \bibnamefont{Pinkse}},
  \bibnamefont{and} \bibinfo{author}{\bibfnamefont{G.}~\bibnamefont{Rempe}},
  \bibinfo{journal}{Phys. Rev. Lett.} \textbf{\bibinfo{volume}{92}},
  \bibinfo{pages}{223001} (\bibinfo{year}{2004}).

\bibitem[{\citenamefont{van Veldhoven et~al.}(2005)\citenamefont{van Veldhoven,
  Bethlem, and Meijer}}]{VelBet05}
\bibinfo{author}{\bibfnamefont{J.}~\bibnamefont{van Veldhoven}},
  \bibinfo{author}{\bibfnamefont{H.~L.} \bibnamefont{Bethlem}},
  \bibnamefont{and} \bibinfo{author}{\bibfnamefont{G.}~\bibnamefont{Meijer}},
  \bibinfo{journal}{Phys. Rev. Lett.} \textbf{\bibinfo{volume}{94}},
  \bibinfo{pages}{083001} (\bibinfo{year}{2005}).

\bibitem[{\citenamefont{Cornell et~al.}(1991)\citenamefont{Cornell, Monroe, and
  Wieman}}]{CorMon91}
\bibinfo{author}{\bibfnamefont{E.~A.} \bibnamefont{Cornell}},
  \bibinfo{author}{\bibfnamefont{C.}~\bibnamefont{Monroe}}, \bibnamefont{and}
  \bibinfo{author}{\bibfnamefont{C.~E.} \bibnamefont{Wieman}},
  \bibinfo{journal}{Phys. Rev. Lett.} \textbf{\bibinfo{volume}{67}},
  \bibinfo{pages}{2439} (\bibinfo{year}{1991}).

\bibitem[{\citenamefont{Kishimoto et~al.}(2006)\citenamefont{Kishimoto,
  Hachisu, Fujiki, Nagato, Yasuda, and Katori}}]{KisHac06}
\bibinfo{author}{\bibfnamefont{T.}~\bibnamefont{Kishimoto}},
  \bibinfo{author}{\bibfnamefont{H.}~\bibnamefont{Hachisu}},
  \bibinfo{author}{\bibfnamefont{J.}~\bibnamefont{Fujiki}},
  \bibinfo{author}{\bibfnamefont{K.}~\bibnamefont{Nagato}},
  \bibinfo{author}{\bibfnamefont{M.}~\bibnamefont{Yasuda}}, \bibnamefont{and}
  \bibinfo{author}{\bibfnamefont{H.}~\bibnamefont{Katori}},
  \bibinfo{journal}{Phys. Rev. Lett.} \textbf{\bibinfo{volume}{96}},
  \bibinfo{pages}{123001} (\bibinfo{year}{2006}).

\bibitem[{\citenamefont{Schlunk et~al.}(2007)\citenamefont{Schlunk, Marian,
  Geng, Mosk, Meijer, and Sch\"{o}llkopf}}]{SchMar07}
\bibinfo{author}{\bibfnamefont{S.}~\bibnamefont{Schlunk}},
  \bibinfo{author}{\bibfnamefont{A.}~\bibnamefont{Marian}},
  \bibinfo{author}{\bibfnamefont{P.}~\bibnamefont{Geng}},
  \bibinfo{author}{\bibfnamefont{A.~P.} \bibnamefont{Mosk}},
  \bibinfo{author}{\bibfnamefont{G.}~\bibnamefont{Meijer}}, \bibnamefont{and}
  \bibinfo{author}{\bibfnamefont{W.}~\bibnamefont{Sch\"{o}llkopf}},
  \bibinfo{journal}{Phys. Rev. Lett.} \textbf{\bibinfo{volume}{98}},
  \bibinfo{pages}{223002} (\bibinfo{year}{2007}).

\bibitem[{\citenamefont{Petrich et~al.}(1995)\citenamefont{Petrich, Anderson,
  Ensher, and Cornell}}]{PetAnd95}
\bibinfo{author}{\bibfnamefont{W.}~\bibnamefont{Petrich}},
  \bibinfo{author}{\bibfnamefont{M.~H.} \bibnamefont{Anderson}},
  \bibinfo{author}{\bibfnamefont{J.~R.} \bibnamefont{Ensher}},
  \bibnamefont{and} \bibinfo{author}{\bibfnamefont{E.~A.}
  \bibnamefont{Cornell}}, \bibinfo{journal}{Phys. Rev. Lett.}
  \textbf{\bibinfo{volume}{74}}, \bibinfo{pages}{3352} (\bibinfo{year}{1995}).

\bibitem[{\citenamefont{Franzosi et~al.}(2004)\citenamefont{Franzosi, Zambon,
  and Arimondo}}]{FraZam04}
\bibinfo{author}{\bibfnamefont{R.}~\bibnamefont{Franzosi}},
  \bibinfo{author}{\bibfnamefont{B.}~\bibnamefont{Zambon}}, \bibnamefont{and}
  \bibinfo{author}{\bibfnamefont{E.}~\bibnamefont{Arimondo}},
  \bibinfo{journal}{Phys. Rev. A} \textbf{\bibinfo{volume}{70}},
  \bibinfo{pages}{053603} (\bibinfo{year}{2004}).

\bibitem[{\citenamefont{Milner et~al.}(2001)\citenamefont{Milner, Hanssen,
  Campbell, and Raizen}}]{MilHan01}
\bibinfo{author}{\bibfnamefont{V.}~\bibnamefont{Milner}},
  \bibinfo{author}{\bibfnamefont{J.~L.} \bibnamefont{Hanssen}},
  \bibinfo{author}{\bibfnamefont{W.~C.} \bibnamefont{Campbell}},
  \bibnamefont{and} \bibinfo{author}{\bibfnamefont{M.~G.}
  \bibnamefont{Raizen}}, \bibinfo{journal}{Phys. Rev. Lett.}
  \textbf{\bibinfo{volume}{86}}, \bibinfo{pages}{1514} (\bibinfo{year}{2001}).

\bibitem[{\citenamefont{Friedman et~al.}(2001)\citenamefont{Friedman, Kaplan,
  Carasso, and Davidson}}]{FriKap01}
\bibinfo{author}{\bibfnamefont{N.}~\bibnamefont{Friedman}},
  \bibinfo{author}{\bibfnamefont{A.}~\bibnamefont{Kaplan}},
  \bibinfo{author}{\bibfnamefont{D.}~\bibnamefont{Carasso}}, \bibnamefont{and}
  \bibinfo{author}{\bibfnamefont{N.}~\bibnamefont{Davidson}},
  \bibinfo{journal}{Phys. Rev. Lett.} \textbf{\bibinfo{volume}{86}},
  \bibinfo{pages}{1518} (\bibinfo{year}{2001}).

\bibitem[{\citenamefont{Ahmadi et~al.}(2005)\citenamefont{Ahmadi, Timmons, and
  Summy}}]{AhmTim05}
\bibinfo{author}{\bibfnamefont{P.}~\bibnamefont{Ahmadi}},
  \bibinfo{author}{\bibfnamefont{B.~P.} \bibnamefont{Timmons}},
  \bibnamefont{and} \bibinfo{author}{\bibfnamefont{G.~S.} \bibnamefont{Summy}},
  \bibinfo{journal}{Phys. Rev. A} \textbf{\bibinfo{volume}{72}},
  \bibinfo{pages}{023411} (\bibinfo{year}{2005}).

\bibitem[{\citenamefont{Schnelle et~al.}(2008)\citenamefont{Schnelle, van
  Ooijen, Davis, Heckenberg, and Rubinsztein-Dunlop}}]{SchOoi08}
\bibinfo{author}{\bibfnamefont{S.~K.} \bibnamefont{Schnelle}},
  \bibinfo{author}{\bibfnamefont{E.~D.} \bibnamefont{van Ooijen}},
  \bibinfo{author}{\bibfnamefont{M.~J.} \bibnamefont{Davis}},
  \bibinfo{author}{\bibfnamefont{N.~R.} \bibnamefont{Heckenberg}},
  \bibnamefont{and}
  \bibinfo{author}{\bibfnamefont{H.}~\bibnamefont{Rubinsztein-Dunlop}},
  \bibinfo{journal}{Opt. Express} \textbf{\bibinfo{volume}{16}},
  \bibinfo{pages}{1405} (\bibinfo{year}{2008}).

\bibitem[{\citenamefont{Sasaki et~al.}(1991)\citenamefont{Sasaki, Koshioka,
  Misawa, Kitamura, and Masuhara}}]{SasKos91}
\bibinfo{author}{\bibfnamefont{K.}~\bibnamefont{Sasaki}},
  \bibinfo{author}{\bibfnamefont{M.}~\bibnamefont{Koshioka}},
  \bibinfo{author}{\bibfnamefont{H.}~\bibnamefont{Misawa}},
  \bibinfo{author}{\bibfnamefont{N.}~\bibnamefont{Kitamura}}, \bibnamefont{and}
  \bibinfo{author}{\bibfnamefont{H.}~\bibnamefont{Masuhara}},
  \bibinfo{journal}{Opt. Lett.} \textbf{\bibinfo{volume}{16}},
  \bibinfo{pages}{1463} (\bibinfo{year}{1991}).

\bibitem[{\citenamefont{Neuman and Block}(2004)}]{NeuBlo04}
\bibinfo{author}{\bibfnamefont{K.~C.} \bibnamefont{Neuman}} \bibnamefont{and}
  \bibinfo{author}{\bibfnamefont{S.~M.} \bibnamefont{Block}},
  \bibinfo{journal}{Rev. Sci. Instrum.} \textbf{\bibinfo{volume}{75}},
  \bibinfo{pages}{2787} (\bibinfo{year}{2004}).

\bibitem[{\citenamefont{Landau and Lifschitz}(1976)}]{LanLif76}
\bibinfo{author}{\bibfnamefont{L.~D.} \bibnamefont{Landau}} \bibnamefont{and}
  \bibinfo{author}{\bibfnamefont{E.~M.} \bibnamefont{Lifschitz}},
  \emph{\bibinfo{title}{Mechanics}} (\bibinfo{publisher}{Pergamon Press},
  \bibinfo{address}{Oxford}, \bibinfo{year}{1976}).

\bibitem[{\citenamefont{Cook et~al.}(1985)\citenamefont{Cook, Shankland, and
  Wells}}]{CooSha85}
\bibinfo{author}{\bibfnamefont{R.~J.} \bibnamefont{Cook}},
  \bibinfo{author}{\bibfnamefont{D.~G.} \bibnamefont{Shankland}},
  \bibnamefont{and} \bibinfo{author}{\bibfnamefont{A.~L.} \bibnamefont{Wells}},
  \bibinfo{journal}{Phys. Rev. A} \textbf{\bibinfo{volume}{31}},
  \bibinfo{pages}{564} (\bibinfo{year}{1985}).

\bibitem[{\citenamefont{Grozdanov and Rakovic}(1988)}]{GroRak88}
\bibinfo{author}{\bibfnamefont{T.~P.} \bibnamefont{Grozdanov}}
  \bibnamefont{and} \bibinfo{author}{\bibfnamefont{M.~J.}
  \bibnamefont{Rakovic}}, \bibinfo{journal}{Phys. Rev. A}
  \textbf{\bibinfo{volume}{38}}, \bibinfo{pages}{1739} (\bibinfo{year}{1988}).

\bibitem[{\citenamefont{Rahav et~al.}(2003)\citenamefont{Rahav, Gilary, and
  Fishman}}]{RahGil03}
\bibinfo{author}{\bibfnamefont{S.}~\bibnamefont{Rahav}},
  \bibinfo{author}{\bibfnamefont{I.}~\bibnamefont{Gilary}}, \bibnamefont{and}
  \bibinfo{author}{\bibfnamefont{S.}~\bibnamefont{Fishman}},
  \bibinfo{journal}{Phys. Rev. Lett.} \textbf{\bibinfo{volume}{91}},
  \bibinfo{pages}{110404} (\bibinfo{year}{2003}).

\bibitem[{\citenamefont{Ridinger and Davidson}(2007)}]{RidDav07}
\bibinfo{author}{\bibfnamefont{A.}~\bibnamefont{Ridinger}} \bibnamefont{and}
  \bibinfo{author}{\bibfnamefont{N.}~\bibnamefont{Davidson}},
  \bibinfo{journal}{Phys. Rev. A} \textbf{\bibinfo{volume}{76}},
  \bibinfo{pages}{013421} (\bibinfo{year}{2007}).

\bibitem[{foo({\natexlab{a}})}]{footnoteARCW1}
\bibinfo{note}{A more rigorous treatment shows that~$F$ is an
  operator~\cite{RahGil03} which can be expanded in powers of $\omega^{-1}$.
  The leading-order term is the function considered here.}

\bibitem[{foo({\natexlab{b}})}]{footnoteARCW2}
\bibinfo{note}{This analogy only approximately holds for the particle's real
  motion, since the particle's micromotion is affected differently by a phase
  hop than by a collision: A phase hop induces a change of phase of the
  oscillating phase factor~$e^{-iF}$ of~$\psi$ and thus instantaneously changes
  the phase of the particle's micromotion, whereas a collision does
  not~\cite{MajDeh67}.}

\bibitem[{foo({\natexlab{c}})}]{footnoteARCW5}
\bibinfo{note}{A stationary mean-motion state obeys the relation
  $\phi_n'(x,t_{\rm ph})^*\phi_n(x,t_{\rm ph})=\phi_n(x,t_{\rm
  ph})^*\phi_n'(x,t_{\rm ph})$, because as an eigenstate of a time-independent
  Hamiltonian it can be written in the form $ \phi_n(x,t)=\xi_n(x)\exp(-iE_n
  t/\hbar). $ Thus, one has $\phi_n'(x,t)=\xi_n'(x)\exp(-iE_n t/\hbar)$ and
  thus $\phi_n'(x,t_{\rm ph})^*\phi_n(x,t_{\rm ph}) = \xi_n'(x) \xi_n(x) =
  \phi_n(x,t_{\rm ph})^*\phi_n'(x,t_{\rm ph})$.}

\bibitem[{\citenamefont{Salomon}(2008)}]{footnoteARCW4}
\bibinfo{author}{\bibfnamefont{C.}~\bibnamefont{Salomon}}
  (\bibinfo{year}{2008}), \bibinfo{note}{private communication}.

\bibitem[{\citenamefont{Thomas et~al.}(2005)\citenamefont{Thomas, Kinast, and
  Turlapov}}]{ThoKin05}
\bibinfo{author}{\bibfnamefont{J.~E.} \bibnamefont{Thomas}},
  \bibinfo{author}{\bibfnamefont{J.}~\bibnamefont{Kinast}}, \bibnamefont{and}
  \bibinfo{author}{\bibfnamefont{A.}~\bibnamefont{Turlapov}},
  \bibinfo{journal}{Phys. Rev. Lett.} \textbf{\bibinfo{volume}{95}},
  \bibinfo{pages}{120402} (\bibinfo{year}{2005}).

\bibitem[{foo({\natexlab{d}})}]{footnoteARCW3}
\bibinfo{note}{In most experimental
  setups~\cite{LeiBla03,MajGhe04,JunRie04,VelBet05,CorMon91,KisHac06,SchMar07,%
PetAnd95,FraZam04,MilHan01,FriKap01,AhmTim05,SchOoi08} it is possible to
  generate such small time-scales, although the period of the ROP is already
  very small.}

\bibitem[{\citenamefont{Grossmann et~al.}(1991)\citenamefont{Grossmann,
  Dittrich, Jung, and {{H\"anggi}}}}]{GrossmannEtAl91}
\bibinfo{author}{\bibfnamefont{F.}~\bibnamefont{Grossmann}},
  \bibinfo{author}{\bibfnamefont{T.}~\bibnamefont{Dittrich}},
  \bibinfo{author}{\bibfnamefont{P.}~\bibnamefont{Jung}}, \bibnamefont{and}
  \bibinfo{author}{\bibfnamefont{P.}~\bibnamefont{{{H\"anggi}}}},
  \bibinfo{journal}{Phys. Rev. Lett.} \textbf{\bibinfo{volume}{67}},
  \bibinfo{pages}{516} (\bibinfo{year}{1991}).

\bibitem[{\citenamefont{Kierig et~al.}(2008)\citenamefont{Kierig,
  Schnorrberger, Schietinger, Tomkovic, and Oberthaler}}]{KierigEtAl08}
\bibinfo{author}{\bibfnamefont{E.}~\bibnamefont{Kierig}},
  \bibinfo{author}{\bibfnamefont{U.}~\bibnamefont{Schnorrberger}},
  \bibinfo{author}{\bibfnamefont{A.}~\bibnamefont{Schietinger}},
  \bibinfo{author}{\bibfnamefont{J.}~\bibnamefont{Tomkovic}}, \bibnamefont{and}
  \bibinfo{author}{\bibfnamefont{M.~K.} \bibnamefont{Oberthaler}},
  \bibinfo{journal}{Phys. Rev. Lett.} \textbf{\bibinfo{volume}{100}},
  \bibinfo{eid}{190405} (\bibinfo{year}{2008}).

\bibitem[{\citenamefont{Eckardt et~al.}(2005)\citenamefont{Eckardt,
  Jinasundera, Weiss, and Holthaus}}]{EckardtEtAl05}
\bibinfo{author}{\bibfnamefont{A.}~\bibnamefont{Eckardt}},
  \bibinfo{author}{\bibfnamefont{T.}~\bibnamefont{Jinasundera}},
  \bibinfo{author}{\bibfnamefont{C.}~\bibnamefont{Weiss}}, \bibnamefont{and}
  \bibinfo{author}{\bibfnamefont{M.}~\bibnamefont{Holthaus}},
  \bibinfo{journal}{Phys. Rev. Lett.} \textbf{\bibinfo{volume}{95}},
  \bibinfo{pages}{200401} (\bibinfo{year}{2005}).

\bibitem[{\citenamefont{Sias et~al.}(2008)\citenamefont{Sias, Lignier, Singh,
  Zenesini, Ciampini, Morsch, and Arimondo}}]{SiasEtAl07}
\bibinfo{author}{\bibfnamefont{C.}~\bibnamefont{Sias}},
  \bibinfo{author}{\bibfnamefont{H.}~\bibnamefont{Lignier}},
  \bibinfo{author}{\bibfnamefont{Y.~P.} \bibnamefont{Singh}},
  \bibinfo{author}{\bibfnamefont{A.}~\bibnamefont{Zenesini}},
  \bibinfo{author}{\bibfnamefont{D.}~\bibnamefont{Ciampini}},
  \bibinfo{author}{\bibfnamefont{O.}~\bibnamefont{Morsch}}, \bibnamefont{and}
  \bibinfo{author}{\bibfnamefont{E.}~\bibnamefont{Arimondo}},
  \bibinfo{journal}{Phys. Rev. Lett.} \textbf{\bibinfo{volume}{100}},
  \bibinfo{pages}{040404} (\bibinfo{year}{2008}).

\bibitem[{\citenamefont{Major and Dehmelt}(1968)}]{MajDeh67}
\bibinfo{author}{\bibfnamefont{F.~G.} \bibnamefont{Major}} \bibnamefont{and}
  \bibinfo{author}{\bibfnamefont{H.~G.} \bibnamefont{Dehmelt}},
  \bibinfo{journal}{Phys. Rev.} \textbf{\bibinfo{volume}{170}},
  \bibinfo{pages}{91} (\bibinfo{year}{1968}).

\end{thebibliography}
\end{document}